\documentclass[%
 reprint,
 amsmath,amssymb,
 aps,
]{revtex4-2}

\usepackage{graphicx}
\usepackage{dcolumn}
\usepackage{bm}
\usepackage[colorlinks,linkcolor=blue,urlcolor=blue,citecolor=blue]{hyperref}

\begin{document}

\preprint{APS/123-QED}

\title{Demonstration of Direct-amplification Enabled Harmonic Generation in an Ultraviolet Free-Electron Laser}



\author{Hao Sun}
\author{Jitao Sun}
\author{Li Zeng}
\author{Yifan Liang}
\author{Lingjun Tu}
\author{Huaiqian Yi}
\author{Qinming Li}
\author{Xiaofan Wang}
\email{wangxf@mail.iasf.ac.cn}
\author{Yong Yu}
\email{yuyong@mail.iasf.ac.cn}
\affiliation{
 Institute of Advanced Light Source Facilities, Shenzhen, Shenzhen 518107, China
}
\author{Jiayue Yang}
\author{Zhigang He}
\author{Yuhuan Tian}
\author{Likai Wang}
\author{Zequn Wang}
\author{Guorong Wu}
\author{Weiqing Zhang}%
 \email{weiqingzhang@dicp.ac.cn}
 \author{Xueming Yang}
\affiliation{
 Dalian Institute of Chemical Physics, Chinese Academy of Sciences, Dalian 116023, China
}

\date{\today}

\begin{abstract}

We report the experimental demonstration of direct-amplification enabled harmonic generation in an ultraviolet free-electron laser (FEL) driven by a low-intensity seed laser. By employing a versatile undulator configuration that enables seed amplification and harmonic generation within a unified setup, we achieved over 100-fold energy gain of the seed and observed exponential growth at the second harmonic. The results demonstrate that a sufficiently long modulator can not only amplify a weak seed but also induce strong energy modulation of the electron beam, enabling efficient harmonic bunching. This method markedly relaxes the power requirements on external seed lasers and presents a viable route toward high-repetition-rate, fully coherent FELs.
\end{abstract} 

\maketitle

High-repetition-rate, fully coherent, free-electron laser (FEL) radiation \cite{McNeil2010,science.1055718} has significant scientific applications in the extreme ultraviolet and X-ray wavelength ranges, such as the fine time-resolved analysis of matter with spectroscopy and photon scattering.
Superconducting linacs enable the delivery of high-repetition-rate electron beams, facilitating high average power output from FELs. Based on superconducting accelerators, FEL in Hamburg (FLASH) \cite{DESY_ackermann2007operation} and European X-ray free-electron laser \cite{Weise2017_european} are now under operation at MHz in
burst mode, while the linac coherent light source II (LCLS-II) \cite{Brachmann2019_lcls2}, the Shanghai high repetition rate
XFEL and extreme light facility (SHINE) \cite{shine_zhao2018} and Shenzhen superconducting soft X-ray free-electron laser (S\textsuperscript{3}FEL) \cite{s3fel_wangtupl043} are aiming to generate MHz FELs in
continuous wave mode.

Most global X-ray FEL facilities employ the self-amplified spontaneous emission (SASE) mechanism \cite{kim1986three}, which is initiated by the stochastic noise in the electron bunch. Therefore, SASE suffers from poor longitudinal coherence and large shot-to-shot fluctuations. The significant phase and intensity fluctuations in the SASE scheme severely constrain applications in X-ray spectroscopy. To overcome these limitations, some seeding techniques like self-seeding  and external seeding have been proposed. Self-seeding schemes can be used to improve temporal coherence, but they still suffer from large shot-to-shot energy fluctuations \cite{self_seeding_amann2012demonstration}. Seeded FELs triggered by stable, coherent external lasers ensure output pulses with high temporal coherence and minimal pulse energy fluctuations \cite{HGHG_yu1991generation,HGHG_yu2002theory,stupakov2009using,xiang2009echo}. This advantage has been theoretically calculated and experimentally verified across spectral ranges from ultraviolet to soft X-ray regions \cite{,PhysRevLett.91.074801,
allaria2012highly_HGHG,EEHG_zhao2012first,xiang2010demonstration_EEHG,hemsing2016echo,ribivc2019coherent_EEHG,Feng:22}.
\begin{figure*}[t] 
\centering
\includegraphics[width=15cm]{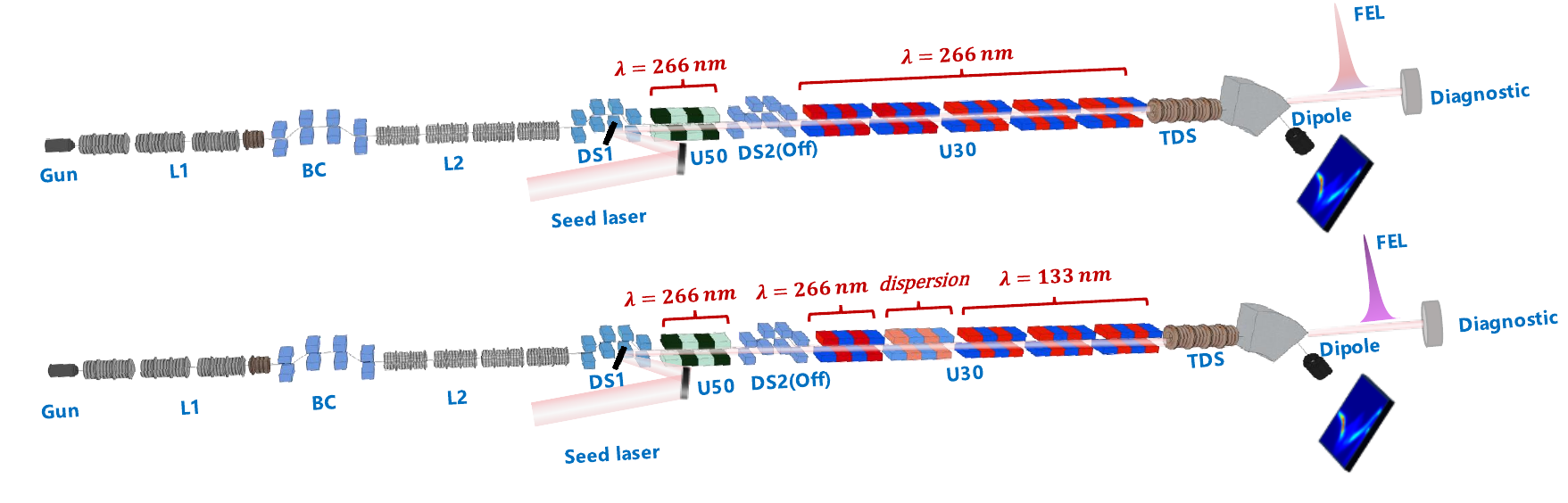}
\caption{ Experimental configurations at the DCLS: (a) Direct seed amplification setup (b) Harmonic generation setup.}
\label{figure1}
\end{figure*}
In seeded FELs like high-gain harmonic generation (HGHG) and echo-enabled harmonic generation (EEHG) that utilize an external laser, the repetition rate is constrained by the seed laser itself. Currently, there are no available seed laser sources that can provide sufficient pulse energy at MHz repetition rates, posing a fundamental challenge for extending externally seeded FELs to high-repetition-rate operation. To address this issue, considerable efforts have been directed toward reducing the power requirements of the seed laser—a key limiting factor. One such approach is the oscillator–amplifier scheme, which incorporates an optical cavity to recirculate the seed laser within the resonator, enabling repeated interactions with successive electron bunches \cite{cavity_paraskaki,cavity_cavity}. In this configuration, the repetition rate is no longer limited by the seed laser but is instead determined by the optical cavity. Another strategy is the optical klystron scheme \cite{yan2021self,OK_GP}, in which a weak seed laser initiates minimal energy modulation, and further modulation is driven by radiation from a weakly prebunched electron beam in an additional section.

Direct seeding was originally proposed as a route toward generating fully coherent FEL pulses by amplifying a seed laser in a long undulator, rather than as a method to reduce seed power requirements, its underlying mechanism lends itself naturally to this goal \cite{DS1,DS2,DS3,DS38nm}. By exploiting the intrinsic amplification capability of direct seeding, the recently proposed direct-amplification enabled harmonic generation (DEHG) scheme \cite{wang2021direct} extends this concept by replacing the short modulator in conventional schemes with a longer one. This allows for simultaneous seed amplification and enhanced energy modulation of the electron beam at the seed laser wavelength, resulting in strong harmonic bunching even from low-power input lasers. Thus, while maintaining the advantages of longitudinal coherence, DEHG effectively lowers the threshold on seed laser power and offers a promising route toward high-repetition-rate, fully coherent FEL operation.

In this Letter, we report on the experimental results of direct amplification and harmonic generation in an ultraviolet free-electron laser driven by a low-intensity laser. The experiment was performed at the Dalian coherent light source (DCLS) \cite{yu:fel19-thp015}, the first high gain FEL user facility in China. A stepwise experimental approach was adopted: an initial setup was configured to demonstrate direct amplification of a low-intensity seed laser, with all undulators tuned to 266 nm [Fig. 1(a)]. Building upon this, the undulator settings were adjusted to enable second-harmonic generation within the same system [Fig. 1(b)]: the U50 and first U30 undulators remained resonant at 266 nm to support seed amplification, while the last three U30 undulators were tuned to 133 nm to produce harmonic radiation. The results from both stages are presented in detail below.

As illustrated in Fig. 1, the DCLS consists  of an injector system, a linear accelerator, a seed laser system, and an undulator system. The electron beam with a bunch charge of 500 pC is generated by a photocathode radio-frequency (RF) gun operating at a frequency of 2856 MHz and a repetition rate of 10 Hz.
It is subsequently accelerated by a linear accelerator that includes three 3-meter-long S-band accelerating structures (L1) and four additional S-band structures (L2), enabling beam energies up to 300 MeV. A magnetic bunch compressor (BC) shortens the electron beam to approximately 2.5 ps in full bunch length, resulting in an average current of 200 A. The normalized transverse emittance is around 1.5 mm · mrad. The undulator system includes a 1-meter-long modulation undulator (U50) with a period of 50 mm, and a radiator section consisting of five 3-meter-long variable-gap undulators (U30) with 30 mm period lengths. An S-band transverse deflecting structure (TDS) is installed downstream for longitudinal phase space diagnostics. 

To enable efficient seeding under improved synchronization and overlap conditions, a dedicated configuration was implemented. The seed laser is generated by the third harmonic of a Ti:Sapphire laser, with a wavelength of 266.7 nm, a pulse width of 1 ps FWHM, an FWHM bandwidth of $5 \times 10^{-4}$, and a peak power of up to 500 MW, replacing the previously used optical parametric amplifier (OPA) source. To enhance spatiotemporal overlap with the electron beam and increase modulation efficiency, the seed laser was focused not in the U50 modulator but near the entrance of the downstream radiator (U30). The mirror used to reflect the laser in DS2 was removed, allowing the seed laser to propagate directly into U30 and interact further with the electron beam. In the experiment, both the U50 and the U30 were tuned to be resonant at the seed laser wavelength of 266.7 nm. As shown in Fig. 2, the overlapping deep-blue region indicates the range of beam energies for which resonance can be simultaneously achieved in both undulators. To operate within this range, the acceleration gradient and RF phase of the accelerating structures were carefully adjusted, yielding a final measured beam energy of 216 MeV—ensuring resonance at the target wavelength and facilitating efficient energy modulation and amplification. Moreover, at this energy, the U30 undulator also supports harmonic radiation at 133 nm, facilitating studies on second-harmonic generation.

\begin{figure} [h]
\includegraphics[width=6.5cm]{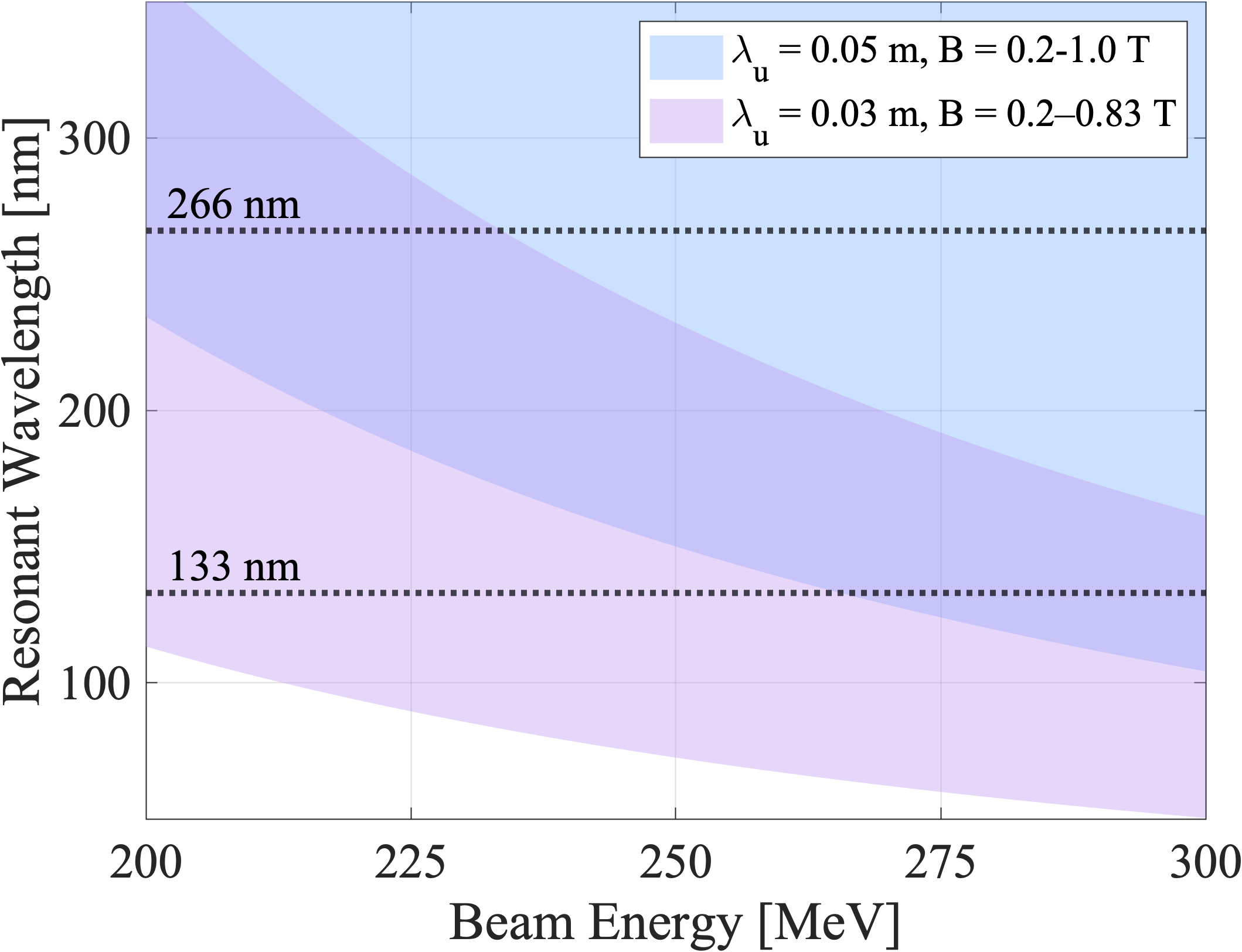}
\caption{Relationship between the FEL resonant wavelength and the electron beam energy for undulators with different period lengths.}
\label{figure0}
\end{figure}

We characterized the RMS slice energy spread of the electron beam as 38 keV using the coherent radiation method \cite{slice_es_measurment}, providing critical input for the subsequent seed laser interaction study. A seed laser with a peak power of 3.85 MW was employed, yielding a power density of $5.37 \times 10^7 \mathrm{~W} / \mathrm{cm}^2$ based on a beam waist radius of 1.5 mm, chosen to fully cover the electron beam. The laser interacted with the electron beam in the modulation undulator (U50), initiating energy modulation. In the downstream radiation undulator (U30), the laser-induced microbunching underwent exponential gain, leading to a significant enhancement of the modulation amplitude and coherent radiation. The electron-laser temporal overlap is ensured by an optical delay line. Temporal synchronization is achieved by monitoring the spontaneous emission of the electron beam and seed laser signals on a fast photodiode located at the exit of the undulator.  For the diagnostics of FEL light, the FEL spectra were obtained using a high-resolution spectrometer (Ocean Optics HR4000), with a spectral coverage from 200 to 400 nm and a resolution of 0.02 nm at 266 nm. 

\begin{figure} [h]
\includegraphics[width=8.5cm]{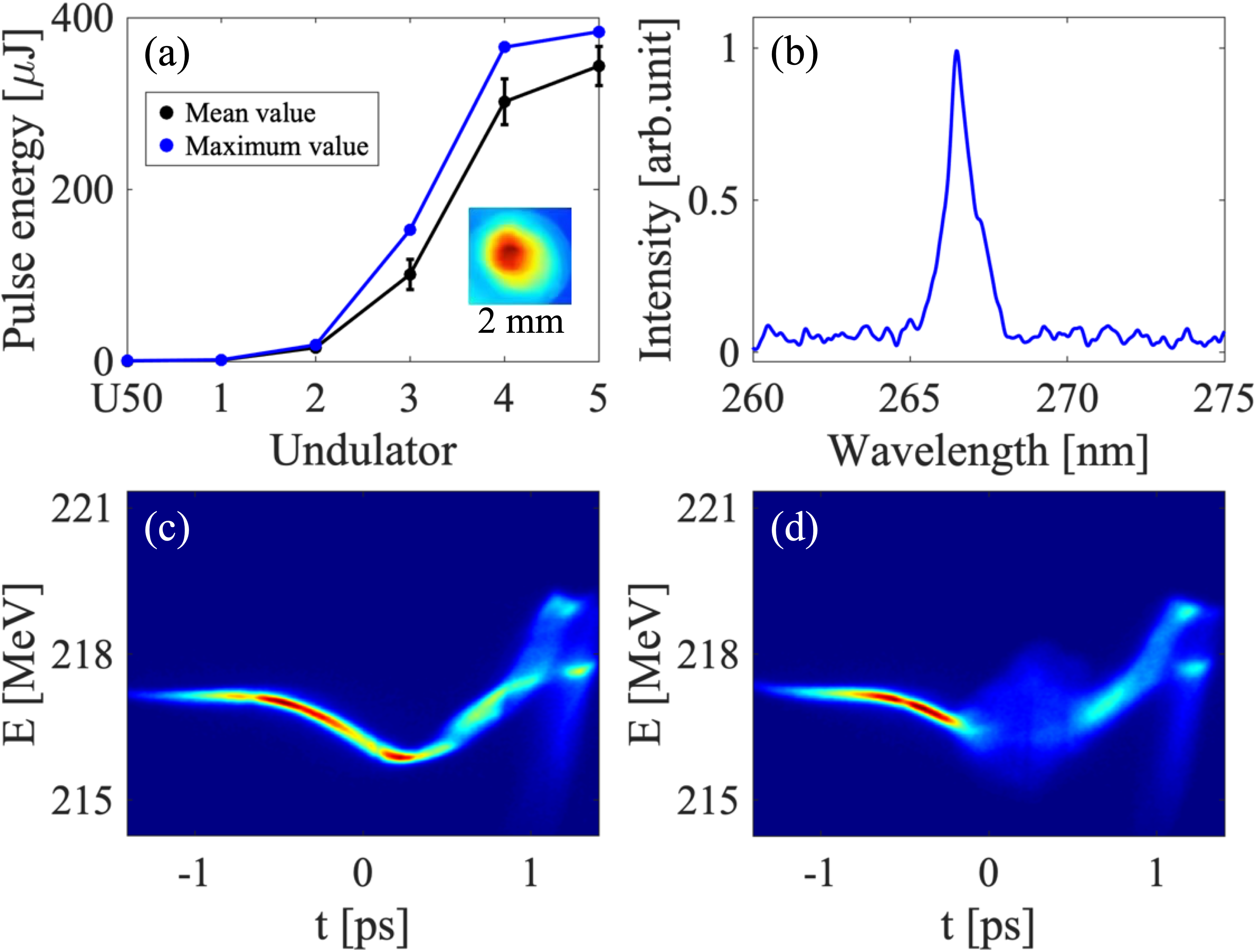}
\caption{Measured FEL gain curves (a) and spectra (b) for direct amplification of deep ultraviolet free-electron laser. (averaged over 100 consecutive shots). The inset in (a) shows measured FEL transverse spot. Measured longitudinal phase spaces of the electron beam by the S-band deflecting cavity at the undulators exit: (c) seed laser off; (d) seed laser on.}
\label{figure3}
\end{figure}

The FEL pulse energy was measured by photodiodes located downstream of the radiator. Figure 3(a) displays the measured gain curves along the radiator as well as a typical transverse distribution of the laser spot. A maximum FEL energy of 383 microjoules was achieved, corresponding to a 100-fold amplification of the seed laser measured at the radiator entrance (3.85 microjoules). To further quantify the FEL gain, the seed laser energy was also measured at the radiator exit in the absence of the electron beam, yielding only 0.089 microjoules. This large difference indicates misalignment between the seed laser and undulator axis-likely due to imperfect beam transport at 216 MeV in the 3 m-long U30 radiator section with 1 m drift spacing, where the FOFO matching condition is not well satisfied. Despite this, FEL amplification remains highly efficient, with a gain exceeding 4300 relative to the residual seed, highlighting the robustness of the seeding scheme.

The FEL spectrum, averaged over 100 consecutive shots, is presented in Fig. 3(b), exhibiting a mean FWHM bandwidth of $2 \times 10^{-3}$, indicative of good longitudinal coherence.  The measured FEL bandwidth is approximately five times broader than the Fourier transform limit. This broadening is attributed to a nonlinear energy chirp in the electron beam, as shown in Fig. 3(c), which arises from radio-frequency curvature and wakefield effects in the accelerating stage \cite{Fengarticle_chirp}.
 Figures 3(c) and 3(d) present two typical longitudinal phase spaces of the electron beams measured by the TDS at the exit of the undulator section with the seed laser off and on, respectively. When the seed laser is on, a clear degradation of beam quality is observed in the central region of the bunch due to energy extraction, corresponding to an FEL lasing region of approximately 1 ps in duration.

To experimentally verify the radiation performance enabled by the DEHG mechanism, coherent harmonic generation was explored under low-intensity seeding conditions. Owing to the limited electron beam energy at DCLS, the study targeted second harmonic generation at 133 nm, using a 266 nm seed laser. The undulator configuration comprised three stages: (1) the U50 and first U30 undulators resonant at 266 nm for seed amplification, (2) the second U30 tuned off-resonance as a dispersive section to enhance microbunching via energy-position correlation, and (3) the final three U30 undulators resonant at the second harmonic for coherent radiation. 
This staged configuration enables not only efficient amplification of the seed laser, but also strong harmonic bunching, which is crucial for DEHG. 
With a seed laser delivering 5.7 MW peak power ($8.06 \times 10^7 \mathrm{~W} / \mathrm{cm}^2$), pulse energies measured 5.7 microjoules at the undulator entrance and 0.13 microjoules at the exit in the absence of the electron beam.


\begin{figure}[htb!] 
\includegraphics[width=8.5cm]{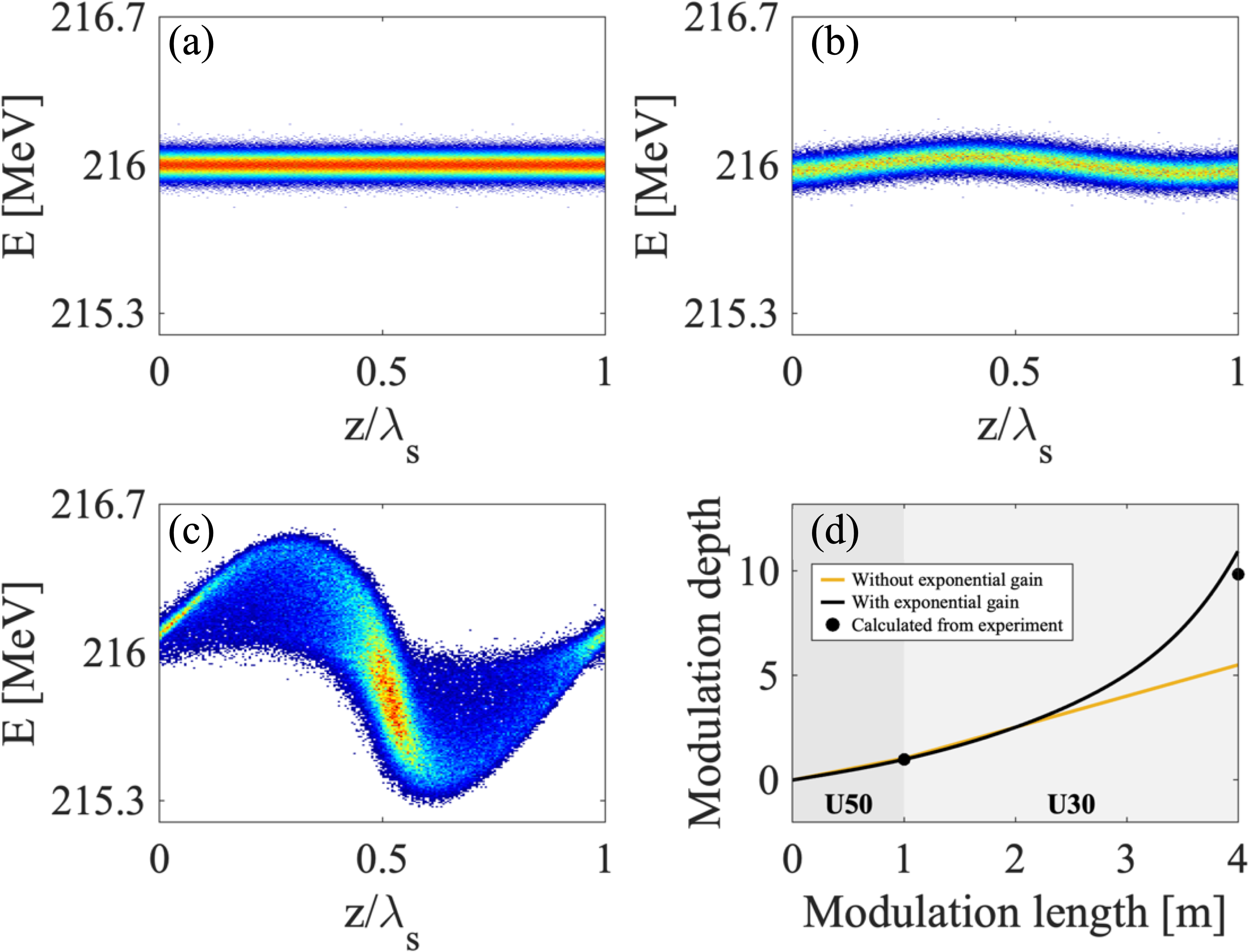}
\caption{Evolution of the electron beam phase space on the wavelength scale of the laser: (a) at the Linac exit, (b) at the U50 undulator exit, and (c) at the first U30 undulator exit. (d) The relationship between modulation depth and modulation segment length.}
\label{figure4}
\end{figure}


A key feature of DEHG lies in its coupled amplification and modulation dynamics: the weak seed laser initiates energy modulation in the U50 modulator, while the first U30 radiator amplifies both the laser field and beam modulation. Figure 4 illustrates this through simulated longitudinal phase space evolution at key undulator locations, resolved at laser wavelength scale. Using Genesis simulations \cite{reiche1999genesis} with experimentally derived parameters, we observed the electron beam acquires an energy modulation of approximately one times the slice energy spread after U50. Continued interaction with the seed laser in the first U30 leads to laser amplification, resulting in a significantly enhanced modulation depth. The simulated modulation reaches nearly 11 times the slice energy spread, which is in reasonable agreement with the value of 9.85 times inferred from the measured radiation energies at the U50 and U30 exits, as shown in Fig. 4(d).

This performance starkly contrasts with non-amplified scenarios, where modulation depths would plateau at 5.5 times the energy spread—demonstrating the crucial role of amplification in boosting modulation. Furthermore, under a typical DCLS configuration where only the 1-meter-long U50 undulator is used for modulation, achieving a comparable energy modulation of 9.85 would require a seed laser peak power as high as 340 MW, far exceeding the 5.7 MW used in this setup. This remarkable efficiency, enabled by the intrinsic coupling between laser amplification and beam modulation, is a hallmark of the DEHG mechanism, fundamentally distinguishing it from conventional seeded FEL schemes, where amplification and modulation are typically treated as sequential and independent processes.

\begin{figure}[h] 
\includegraphics[width=8.5cm]{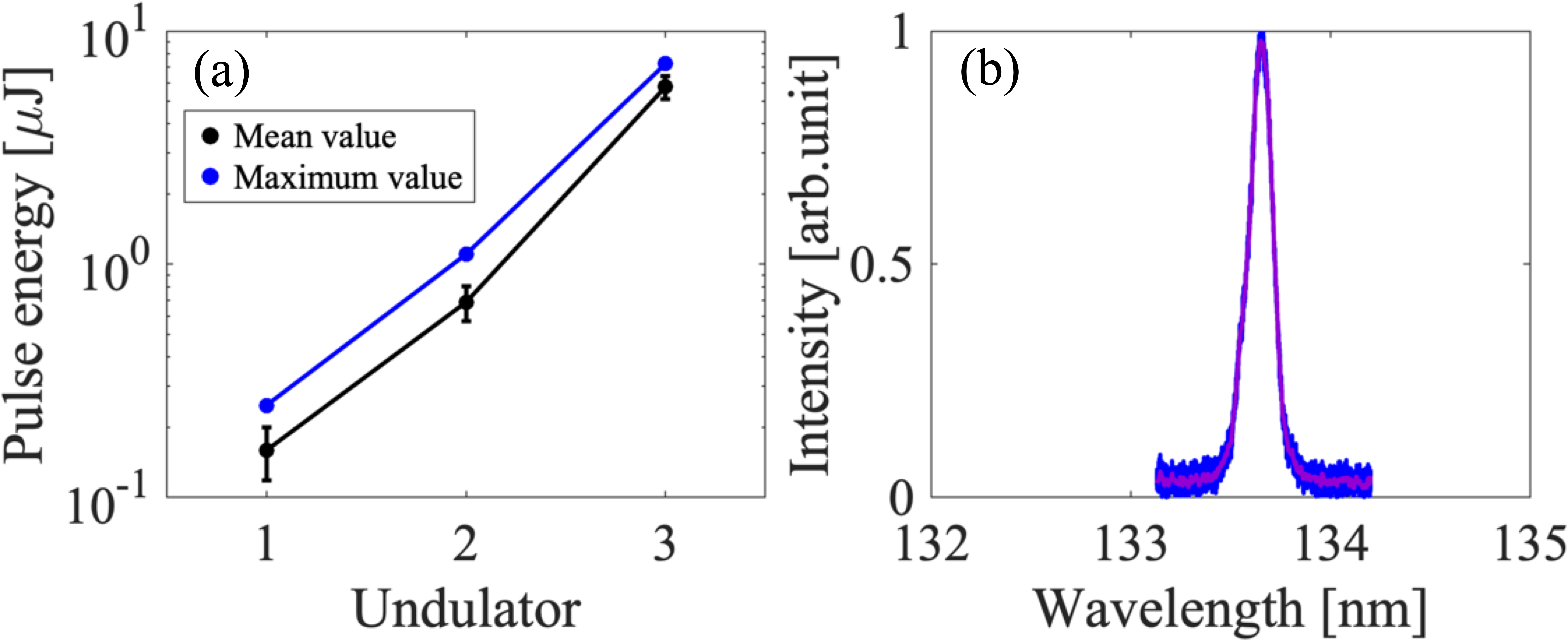}
\caption{Measured FEL gain curves (a) along the last three undulators and measured spectra (b) for harmonic generation of a vacuum ultraviolet free-electron laser. (averaged over 1000 consecutive shots). }
\label{figure5}
\end{figure}

The FEL output was characterized using an online spectrometer spanning 50–150 nm with 0.03 nm resolution at 133 nm \cite{spectrometer_Wang}, and pulse energy was monitored by calibrated photodiodes downstream of the radiator. 
Figure 5(a) displays the measured FEL gain curves along the last three U30 undulators, highlighting a clear exponential amplification process that drives the second harmonic radiation to a peak pulse energy of 7 microjoules. The FEL spectrum, averaged over 1000 consecutive shots, is displayed in Fig. 5(b) and features a narrow FWHM bandwidth of $1 \times 10^{-3}$, demonstrating excellent longitudinal coherence inherited from the seed laser. The maximum achievable pulse energy in this setup is constrained by several practical factors: the relatively low electron beam energy limits the establishment of an optimal transverse focusing structure within the 3-meter undulators \cite{DS3}; the beam trajectory is highly sensitive to undulator field integrals, complicating precise orbit control; and the finite number of radiator segments restricts the overall radiation efficiency. Despite these constraints, the experiment clearly demonstrates the essential features of the DEHG mechanism—namely, efficient harmonic generation, strong bunching, and preserved temporal coherence.

In summary, we have experimentally demonstrated the direct amplification and harmonic generation in an ultraviolet free-electron laser driven by a low-intensity seed laser. A 100-fold amplification of the seed laser energy was successfully achieved under low-intensity condition, and harmonic generation amplification was observed. The stepwise experiment demonstrates that a long modulator facilitates the amplification of a low-intensity laser and enhances the energy modulation of the electron beam at the seed laser wavelength scale to effectively generate harmonic radiation. For a high repetition rate FEL, the peak power required for a seeded FEL decreases from hundreds of MW to a few MW, allowing the laser repetition rate to be increased from 10 kHz to 1 MHz. This method can effectively alleviate the demand for peak power of the seed laser in high-repetition-rate seeded FELs, providing a reliable technical pathway for high-repetition-rate, fully coherent FELs.

We would like to acknowledge the support of the DCLS (https://cstr.cn/31127.02.DCLS). This work is supported by the Scientific Instrument Developing Project of Chinese Academy of Sciences (Grant No. GJJSTD20220001) and the National Natural Science Foundation of China (Grant No. 22288201).

\nocite{*}

\end{document}